\documentclass{amsart}

\usepackage{amsmath}
\usepackage{amssymb}
\usepackage{amsfonts}
\usepackage{amsthm}

\DeclareMathOperator{\diverg}{div}
\DeclareMathOperator{\id}{id}

\DeclareMathOperator{\ran}{ran}
\DeclareMathOperator{\meas}{meas}

\newtheorem{lemma}{Lemma}[section]
\newtheorem{theorem}[lemma]{Theorem}
\newtheorem{proposition}[lemma]{Proposition}
\newtheorem{assumption}[lemma]{Assumption}
\newtheorem{corollary}[lemma]{Corollary}
\theoremstyle{definition}
\newtheorem{definition}[lemma]{Definition}
\theoremstyle{definition}
\newtheorem{remark}[lemma]{Remark}
\theoremstyle{definition}

{\catcode `\@=11 \global\let\AddToReset=\@addtoreset}
\AddToReset{equation}{section}

\newcommand{\N}{{\mathbb N }}
\newcommand{\R}{{\mathbb R}}
\newcommand{\C}{{\mathbb C}}
\newcommand{\e}{{\varepsilon }}
\newcommand{\ie}{{\sl i.e.\/ }}
\newcommand{\cf}{{\sl cf.\/ }}
\newcommand{\eg}{{\sl e.g.\/}}

\def\d{{\partial}}
\def\S{{\mathcal S}}
\def\v{{\tt v}}

\def\({\left(}
\def\){\right)}
\def\<{\left\langle}
\def\>{\right\rangle}
\def\O{\mathcal O}

\newcommand{\newpar}{\par}\parindent =0pt\parskip=3pt\textheight = 615pt

\newcommand{\Id}[1]{{\rm I\kern-2pt I_{#1}}}
\renewcommand{\hbar}{{\displaystyle\bar{\phantom{x}}\kern-6pt h}}

\numberwithin{equation}{section}

\begin{document}


\title[Nonlinear effective mass theorems]{Effective mass theorems for nonlinear Schr\"odinger equations} 

\author[C. Sparber]{Christof Sparber}
\address[C. Sparber]{Institut f\"ur Mathematik der
Universit\"at Wien\\ Nordbergstra\ss e 15\\ A-1090 Vienna\\ Austria}
\email{christof.sparber@univie.ac.at}
\begin{abstract}
We consider time-dependent nonlinear Schr\"odinger equations subject to smooth, lattice-periodic potentials plus additional 
confining potentials, slowly varying on the lattice scale. After an appropriate scaling we study the homogenization limit for 
vanishing lattice spacing. Assuming well prepared initial data, the resulting effective dynamics is governed by a homogenized 
nonlinear Schr\"odinger equation with an effective mass tensor depending on the initially chosen Bloch eigenvalue. 
The given results rigorously 
justify the use of the effective mass formalism for the description of Bose-Einstein condensates on optical lattices.    
\end{abstract}
\thanks{This work has been partially supported by the APART grant of C.S. (funded by the 
Austrian Academy of Science), the Wittgenstein Award 2000 of Peter Markowich 
(funded by the Austrian research fund FWF), and the Warwick EPSRC symposium 
on the ``Mathematics of quantum systems''.}
\maketitle

\begin{center}
version: \today
\end{center}


\section{Introduction}

Recent experiments on \emph{Bose-Einstein condensates} (BECs) study the influence of \emph{optical lattices} (or super-lattices) on the 
dynamics of the condensate, \cf \cite{ChNi, DFK, KMPS, MoAr}. The theoretical description of such systems is usually based on the 
famous \emph{Gross-Pitaevskii equation}, \ie 
\begin{equation}
\label{gpe}
i \hbar \partial _t \psi =  -\frac{\hbar^2}{2m} \Delta \psi +
V(x)\psi + U_0(x)\psi + N \alpha_0 |\psi|^2 \psi,\quad x \in \R^3, \, t\in \R, \\
\end{equation} 
where $m$ is the atomic mass, $\hbar$ is the Planck constant, 
$N$ is the number of atoms in the condensate and 
\begin{equation} 
\alpha_0 = \frac{4\pi\hbar^2 a}{m},
\end{equation}
with $a \in \R$ being the $s$-wave scattering length derived from the corresponding $N$-particle theory, \cf \cite{PiSt}. 
Depending on the sign of $a$, the condensate is said to be either \emph{repulsive} (stable) or \emph{attractive} (unstable). 
In the nonlinear Schr\"odinger equation (NLS) \eqref{gpe} the potential $U_0(x)$ models some given external \emph{confinement}, 
whereas $V(x)$ represents the \emph{lattice-potential}, satisfying
\begin{equation}
\label{Vper}
V(x + \gamma) = V(x), \quad \forall x \in \R^3,
\gamma \in \underline \Gamma, 
\end{equation}
where $\underline \Gamma \simeq \mathbb Z^3$ denotes some given \emph{regular lattice}, generated through a basis
$\{\underline \zeta_1,\underline \zeta_2,\underline \zeta_3\}$, $\underline \zeta_l \in \R^3$, \ie   
\begin{equation}
\label{eq:net}
\underline \Gamma=\left\{\gamma \in \R^3: \ \gamma=\sum_{l=1}^3 \gamma_l \underline \zeta_l, 
\ \gamma_l\in \mathbb Z \right\}.
\end{equation}
Of course the nonlinear dynamics described by \eqref{gpe} can be highly involved. In the physics literature it is therefore frequently proposed, 
\cf \cite{KoSa, PBZBM, StZh}, to consider the following simplifications: First it is assumed that the matter wave-field $\psi(t)$ can be 
characterized by a (fixed) central wave vector $k_0\in \R^3$ and second one tries to capture the rapid oscillations in the wave function $\psi(t)$ 
by performing an asymptotic expansion in terms of \emph{Bloch waves} $\chi_n(y,k_0)$ 
(see equation \eqref{bloch} below for their precise definition). The center of mass of the wave function is then
described by a slowly varying envelope function $f(t,x)$, the dynamics of which is \emph{formally} found to be governed by an 
\emph{effective-mass NLS}. These types of approximations are well known in solid-state physics \cite{Ca}, though mostly in 
a time-independent setting \cite{Pe}, where one considers the motion of electrons in a crystal. It is the purpose of this work to 
\emph{rigorously justify} the described approach in particular within the considered nonlinear context.
\newpar
To this end we shall consider a more general NLS than originally proposed in \eqref{gpe}, namely 
\begin{equation}
\label{nls}
\left \{
\begin{aligned}
i \hbar \partial _t \psi = & \ -\frac{\hbar^2}{2m} \Delta \psi +
V(x)\psi + U_0(t,x)\psi + \alpha |\psi|^{2\sigma} \psi, \quad x \in \R^d, \, t\in \R,\\
\psi \big |_{t=0}   = &  \ \psi_I(x),
\end{aligned}
\right.
\end{equation}
where $\alpha \in \R$, and $ \sigma \in \N$. To motivate the choice $\sigma \geq 1$, we
note that for $d<3$ higher order nonlinearities are frequently used in the description of BECs, \cf \cite{KNSQ, LSY}. 
Moreover different NLS type models of the form \eqref{nls} also appear in nonlinear optics and laser physics, \cf \cite{SuSu} 
(see also \cite{JMR} for a rigorous derivation). 
We assume $\psi_I\in L^2(\R^d)$ to be normalized such that  
\begin{equation}
\label{nor}
\int_{\R^d} |\psi_I(x) |^2 dx =1.
\end{equation}
This normalization condition is henceforth conserved during the time-evolution. Again $V$ is assumed to be periodic w.r.t. 
to some regular lattice $\underline \Gamma \simeq \mathbb Z^d$ and $U_0$ denotes some, in general time-dependent, 
smooth external potential. 
Now, we rescale the equation \eqref{nls}, in order to precisely identify the asymptotic regime we shall be dealing with in 
the following. We have in mind a situation where the potential $U_0$ is slowly varying on the lattice-scale corresponding to $V$. 
Hence, there are essentially two scales in this problem. First, the \emph{macroscopic length-} and \emph{time-scale}, denoted $L$ and 
$T$ respectively, which are introduced via $U_0$ by rewriting it in the following dimensionless form
\begin{equation}
U_0(t,x) = \frac{m L^2}{T^2} \, U(t/T, x/L).
\end{equation} 
In other words, the scaled potential $U$ is such that a free particle of mass $m$ under the influence of $U_0$ will travel the 
distance $L$ in the time unit $T$. On the other hand we can also introduce a couple of \emph{microscopic scales}, $\lambda$ and $\tau$, via a 
rescaling of the periodic potential $V$ such that 
\begin{equation}
V(x) = \frac{m \lambda^2}{\tau^2} \, V_{\Gamma}(x/\lambda).
\end{equation} 
The \emph{rescaled lattice} $\Gamma$ is henceforth generated through a basis $\{\zeta_l\}_{l=1}^3$, 
where $\zeta_l = \underline \zeta_l/\lambda$, 
and the microscopic time-unit $\tau$ is then given by 
\begin{equation}
\label{tau}
\tau = \frac{m \lambda^2}{\hbar}.
\end{equation}
We consequently define two \emph{small dimensionless parameters}, $\e$ and $\delta$ respectively, as being the length- and time-ratios, \ie 
\begin{equation}
\label{epde}
\e = \frac{\lambda}{L}, \quad \delta = \frac{\tau}{T}.
\end{equation}
In the following, both of them are assumed to be \emph{small}, \ie $\e\ll1$, $\delta \ll1$, but in general \emph{not necessarily equal}. 
Next we introduce new space- and time-variables $\tilde x$ and $\tilde t$ via
\begin{equation}
\tilde x = \frac{x}{L}, \quad \tilde t = \frac{t}{T},
\end{equation}
and rescale the NLS \eqref{nls} in dimensionless form. Having in mind the normalization condition \eqref{nor} we also need to rescale the 
wave function $\psi$ by 
\begin{equation}
\tilde\psi(\tilde t, \tilde x) = L^{d/2} \psi(t,x). 
\end{equation}
After multiplying \eqref{nls} by $T^2/(m L^2)$, we consequently arrive at the following dimensionless two-parameter model 
(where we again omit all "$ \ \tilde { } \ $" for simplicity):
\begin{equation}
\label{snls}
\left \{
\begin{aligned}
i h \partial _t \psi = & \  -\frac{h^2}{2} \, \Delta \psi +
\frac{h^2}{\e^2} V_{\Gamma}\left(\frac{x}{\e} \right) \psi + U(t,x)\psi + 
\kappa \, |\psi|^{2\sigma} \psi,\\
\psi \big |_{t=0}   = &  \ \psi_I(x).
\end{aligned}
\right.
\end{equation}
Here we introduced two additional dimensionless parameters
\begin{equation}
\label{hk}
h := \frac{\hbar T}{m L^2}, \quad \kappa := \frac{\alpha T^2}{m L^{d\sigma+2}},
\end{equation}
the former of which can be considered to be Planck's constant in the macroscopic variables. Note that the following 
important relation holds
\begin{equation}
\label{rel}
\e^2  = h \delta ,
\end{equation}
connecting the ratio of the length-scales $\e $ with the corresponding time-scale ratio $\delta $. Finally, since we are aiming 
for nonlinearities of order $\O(1)$ we shall impose from now on that 
\begin{equation}
\label{nlcon}
|\kappa | \equiv \frac{|\alpha| T^2}{m L^{d\sigma+2}} = 1, \ \mbox{or, equivalently,} \  T = \sqrt{\frac{m L^{d\sigma +2}}{|\alpha|}},
\end{equation}
hence relating the macroscopic length- and time-scales in a specific way. We remark that 
in the linear case a scaling analogous to \eqref{snls} has been introduced in \cite{PoRi}.
\newpar
A brief discussion on several aspects of this scaling procedure is now in order: First note that if we choose 
$h = \e$, hence, in view of \eqref{rel}, $\e = \delta$, \ie if we choose the \emph{same} ratio for both, 
the length- and the time-scales, then the equation \eqref{snls} simplifies to a one-parameter model given by
\begin{equation}
\label{scnls}
i \e \partial _t \psi =  \ -\frac{\e^2}{2} \, \Delta \psi + V_{\Gamma}\left(\frac{x}{\e}\right) \psi + 
U(t,x)\psi + \kappa \, |\psi|^{2\sigma} \psi.
\end{equation}
This is nothing but the standard \emph{semi-classical scaling} for (nonlinear) Schr\"odinger-type equations including an additional 
\emph{highly oscillatory periodic potential} $V_\Gamma$. Recently the rigorous study of the corresponding asymptotic regime 
$\e \rightarrow 0$, known as the \emph{combined semi-classical and adiabatic approximation}, attracted lots of interest. 
In particular in the linear setting, \ie $\kappa = 0$, different mathematical approaches 
are currently at hand, \eg, \emph{WKB-type expansions} \cite{BLP, GRT}, \emph{Wigner transformation techniques} \cite{MMP, GMMP}, 
and the so-called \emph{space-adiabatic perturbation theory} \cite{PST, Te}, which gives the most sophisticated mathematical results so far. 
Including nonlinear effects, the literature is not so abundant. To the author's knowledge the only result in this direction is a recent paper 
by R. Carles, P. Markowich, and the author himself \cite{CMS}. The results given there though are only valid for weak nonlinearities, in the 
sense that we need to assume $\kappa \sim \O(\e)$. Therefore a different rescaling of the original NLS \eqref{nls} 
has been introduced in \cite{CMS}.
\begin{remark}
Additionally there exists a related work on the semi-classical limit of the so-called Schr\"odinger-Poisson 
system \cite{BMP} in a crystal. There however additional assumptions have to be imposed which are out of the realm of the 
present setting (like truly mixed-state initial data). 
\end{remark}
In the following our focus is not on the semi-classical regime though. Rather we shall study the asymptotic behavior of the 
scaled NLS \eqref{snls} for $\e \ll 1$ but with a fixed $h$ of order $\O(1)$. 
Note that, by \eqref{rel}, this implies $\delta \sim \O(\e^2)$, hence we are considering our 
system on a much larger macroscopic time-scale $T$ than we do by fixing a macroscopic length-scale $L$ via \eqref{epde}. In particular we are dealing with 
much larger times $T$ as in the semi-classical studies described above. Roughly speaking the semi-classical regime can be seen to 
be an asymptotic description for \emph{ballistic scales}, whereas we shall be dealing in the following with \emph{dispersive scales} 
(sometimes also called \emph{diffusive scales}). As we shall see, this indeed turns out to be the asymptotic regime where one can rigorously justify the effective-mass formalism discussed at the beginning. Remark however that, in contrast to what is noted in \cite{AlPi}, 
the considered regime is \emph{not} equivalent to the one obtained after rescaling time in the semi-classical equation \eqref{scnls} 
by $t \rightarrow \e t$. The reason being the different orders of magnitude in the external potential $U$ and 
in the nonlinearity. 
\newpar
To have a more concrete feeling of the involved time- and length-scales we come back to our original equation \eqref{gpe}. 
Thus we consider \eqref{snls} in $d=3$, with $\sigma =1$ and therefore $\kappa =  4\pi \hbar^2a T^2 /(m L^5)$ by \eqref{hk}. 
A particular example for the periodic potentials used in 
physical experiments is then given by \cite{DFK, PiSt}
\begin{equation}
\label{lasp} 
V\left(x \right)= \sum_{l=1}^3 \frac{\hbar^2 \xi_l^2}{m} \sin^2\left(\xi_{l} x_l \right),\quad \xi_l\in \R, 
\end{equation}
where $\xi=(\xi_1, \xi_2,\xi_3)$ denotes the wave vector of the laser field which generates the optical lattice. 
Hence we readily identify $\lambda = (\lambda_1,\lambda_2,\lambda_3)$ as $\lambda_l = 1/\xi_l$. Moreover 
the slowly varying external potential $U_0$ is usually modeled to be static and of harmonic oscillator type (isotropic or 
anisotropic), \ie 
\begin{equation}
U_0(x) = \frac{m\omega_0^2}{2} \, |x|^2, \quad \omega_0\in \R, \, x\in \R^3.
\end{equation}
In this case, a natural choice for the macroscopic length -scale is therefore given by $L=a_0$, where $a_0$ denotes the length of the harmonic 
oscillator ground state corresponding to $U_0(x)$, \ie 
\begin{equation}
a_0:= \sqrt{\frac{\hbar}{\omega_0 m}}.
\end{equation}
The assumption $\e \ll 1$ is then of course equivalent to $\lambda \ll a_0$. In an actual physical experiment this requirement can be easily 
satisfied as a typical ground state length would be $a_0 \approx 10^{-6} [m]$, whereas the wave vectors $\xi_l$ of the laser fields are 
usually tuned from $10^{6} [1/m]$ to $10^{9} [1/m]$, the latter case therefore being suitable in our situation. 
The corresponding relation for the time-scales though is more subtle. 
From the condition \eqref{nlcon} we see that $T$ has to be chosen such
$$
T^2 = \frac{a_0^5 m^2}{4\pi N |a| \hbar^2} \gg \tau ^2, \ \mbox{since $\delta \ll 1$.}
$$
With $\tau$ given by \eqref{tau} this finally leads to the requirement
$$
\frac{a_0^5}{4\pi N |a|} \gg \lambda ^4.
$$
In particular, in the so-called \emph{moderate interaction regime}, characterized by the fact that $4\pi N |a| \approx a_0$ \cite{BJM}, this 
is again equivalent to $a_0 \gg \lambda$ and in this case we compute  
$$
T^2 = \frac{a_0}{4\pi N|a| \omega_0^2} \approx \frac{1}{\omega_0^2}.
$$
Note that this is precisely what one would get in the corresponding linear situation.
\newpar
From a mathematical point of view the limit $\e \rightarrow 0$ with $\delta$ fixed, corresponds to the so-called \emph{homogenization limit} of 
\eqref{snls}. In view of the classical homogenization results as described in, \eg \, \cite{BLP,JiKo}, the main new difficulty, apart from the 
appearing nonlinearity, stems from the \emph{large factor} $1/\e^2$ in front of the periodic potential which furnishes a highly 
\emph{singularly perturbed} term. It is therefore not a surprise that, even in the linear case, this type 
of homogenization problems have been rigorously studied only very recently \cite{ACPSV}. In particular, (linear) time-dependent Schr\"odinger-type 
equations have been considered in \cite{AlPi} and in \cite{PoRi}. The latter result relies on the use of \emph{Wigner measures}, a technique which 
can not be applied in the given nonlinear situation though. The former work is more closely related to ours, as it combines 
classical homogenization techniques, most notably \emph{two-scale convergence methods} \cite{Al}, with Bloch wave decomposition \cite{CoVa}. 
However, we want to stress the fact that the nonlinear case we shall be dealing with, is by 
no means a straightforward generalization of the linear results.   
More precisely, one should note that the scaling of \eqref{snls} in general prohibits the derivation of suitable, \ie \emph{uniformly} in $\e$, 
a-priori estimates, except for the basic $L^2$ estimate, corresponding to the conservation of mass.  
In the majority of cases, the derivation of such uniform estimates is crucial to gain sufficient control on the limiting behavior 
of the appearing nonlinearities, a problem which can not be handled by using weak-convergence methods (such as Wigner measures or 
two-scale convergence).
\begin{remark} Also, by the same reasons, our results do not fit in the framework of \emph{H-measures} \cite{Ta}, 
or \emph{G-convergence} \cite{Pa}. 
\end{remark}
We note that in \cite{ACPSV,AlPi} the authors propose the usage of a \emph{factorization principle} in order to extend their results also to 
the nonlinear case. This approach though remains unproven there and moreover it is known to be applicable only in situations 
where the initial data $\psi_I$ is \emph{concentrated at the minimum of the first Bloch band} 
(see \cite{ACPSV, AlPi} for more details on this). 
\newpar
In comparison to that, the results given below are indeed \emph{independent} of the number of the Bloch band 
and also they do \emph{not} require $\psi_I$ to be concentrated at a local minimum of the considered band.  
On the other hand we do need the initial data $\psi_I$ to be \emph{well-prepared} in a sense to be made more precise below, \cf 
Assumption \ref{assw}. Additionally we need to assume sufficient regularity on the potentials $U,V$ as well as on 
the initial data $\psi_I$. The reason for these assumptions is on the one hand the fact that we shall use a more 
traditional \emph{multiple-scales expansion method}, similar to those introduced in \cite{BLP}. 
This approach will allow us to obtain, in a rather transparent way, an asymptotic description of 
$\psi(t)$ for small $\e\ll 1$, and also to determine the corresponding \emph{effective homogenized NLS}. 
On the other hand, in order to prove that the given asymptotic solution is indeed \emph{stable} under the nonlinear time-evolution 
governed by \eqref{snls}, we shall adapt an approach originally introduced to prove the accuracy of \emph{nonlinear geometrical optics 
expansions}, \cf \cite{DoRa, Gu, Ra} for the most closely related results to the present work. 
The given proof will then again heavily rely on the fact that we have sufficient regularity 
properties and well prepared initial data. 
A similar strategy recently proved to be successful when applied to the weakly-nonlinear semi-classical situation studied in \cite{CMS}. 
The main goal of this paper though is not the introduction of new methods but rather a complete and correct 
treatment of the problem at hand. Moreover one should keep in mind that for completely arbitrary initial data $\psi_I\in L^2(\R^d)$, 
one can not expect an effective mass type dynamics to be valid. In other words, to obtain an equation of the form \eqref{ineff} the 
initial data $\psi_I$ needs to be (asymptotically) of the same type as stated below, at least in leading order, and the 
additional well-preparedness assumptions we shall need, only concern the higher order terms within the asymptotic expansion. 
Remark that only in linear cases the superposition 
principle allows for more general states $\psi_I$. 
\newpar
On the expense of not completely well defined assumptions (to be made precise later on) let us now state the 
typical \emph{effective mass theorem} we shall prove in the following:
\begin{theorem}\label{theo:typique}
Let $V_\Gamma$ and $U$ be smooth, real-valued potentials, such that $V_\Gamma$ is $\Gamma$-periodic and 
$U$ is sub-quadratic. Assume that for some $k_0 \in \R$, the initial data $\psi_I$ is of the following form 
\begin{equation*}
\psi_I (x) =  f_I(x) \chi_n\left(\frac{x}{\e}, k_0\right) e^{i k_0\cdot x/\e}+\e \eta^\e(x),
\end{equation*}
where $f_I\in \S(\R^d;\C)$, $\chi_n(y,k)$ is an eigenfunction of Bloch's spectral cell problem, corresponding to a 
simple eigenvalue $E_n(k)$, and the corrector 
$\eta^\e\sim \O(1)$ is such that $\psi_I$ is well prepared, up to sufficiently large $K\in \N$, in the sense of Assumption~\ref{assw} below.  
Then there exists a $\e_0>0$, such that for $\e\leq\e_0$ the following asymptotic estimate holds 
\begin{align}\label{es1}
\sup_{t\in[-\tau_0, \tau_0]}
\left\| \psi (t) - v_0^\e(t)
\right\|_{L^2(\R^d)}&=\O(\e) , \quad \tau_0 < \tau,
\end{align}
where $\tau>0$ is the maximum existence-time for a smooth solution $f(t,x)$ for the homogenized NLS 
\begin{equation}\label{ineff}
i h \partial _t f = \,  -\frac{h^2}{2} \, \diverg(M^*\nabla) f + U(x)\psi + 
\kappa^* \, |f|^{2\sigma} f,
\end{equation}
with effective mass tensor $M^*=D^2E_n(k_0)$ and an effective coupling $\kappa^*\in \R$, 
given by \eqref{k} below. 
Moreover, the leading order approximate solution $v_0^\e$ is found to be
\begin{equation}
\label{apso}
v_0^\e(t,x)= f\left(t,x-  \frac{h t}{\e} \nabla_k E_n(k_0) \right ) \chi_n\left(\frac{x}{\e}, k_0\right) e^{i k_0\cdot x/\e}e^ {-i h E_n(k_0)t/\e^2}.
\end{equation}
\end{theorem}
Remark that the estimate \eqref{es1} implies a strong two-scale convergence statement as defined in \cite{Al, ACPSV, AlPi}. 
Also note that, apart from the nonlinearity, our approach represents a refinement of the classical multiple-scales expansions   
given in \cite{BLP}, in the sense that we can include possibly appearing \emph{large drifts} (clearly visible in \eqref{apso} 
in the second argument of $f$) in order to resolve 
the underlying dispersive behavior. 
The possibility of large drifts present in the asymptotic solution can be seen as the aftermath of the ballistic regime 
known from the semi-classical situation, a fact which has already been noticed in linear situations \cite{AlPi, Ga}. 
\newpar
The paper is now organized as follows: In Section \ref{spre} we collect some preliminary results and important notations used 
throughout this work. We then proceed similar to \cite{CMS} and first present in Section \ref{sexp} the multiple scales expansion 
method, whereas its nonlinear stability shall be proved in Section \eqref{sstab}.

 
\section{Preliminaries}\label{spre}

For simplicity we restrict ourselves in this work to \emph{static external potentials} $U=U(x)$, although all results 
could be generalized to the case of time-dependent potentials $U(t,x)$ which are smooth w.r.t. $t\in \R$ (indeed we could as 
well include smoothly time-dependent coupling factors $\kappa(t)\in \R$). Thus we 
study in the following the asymptotic behavior as $\e \rightarrow 0$ of 
\begin{equation}
\label{snls1}
\left \{
\begin{aligned}
i h \partial _t \psi = & \  -\frac{h^2}{2} \, \Delta \psi +
\frac{h^2}{\e^2} V_{\Gamma}\left(\frac{x}{\e} \right) \psi + U(x)\psi + 
\kappa \, |\psi|^{2\sigma} \psi,\\
\psi \big |_{t=0}   = &  \ \psi_I(x).
\end{aligned}
\right.
\end{equation}
All results given below are then valid for potentials which satisfy the following basic assumption.

\begin{assumption}\label{aspot} 
The potentials $U$ and $V_\Gamma$ are such that \ie $V_\Gamma, U \in
C^\infty(\R^d;\R)$, and moreover they satisfy: 
\begin{itemize}
\item[\textbf{(i)}] $V_\Gamma$ is $\Gamma$-periodic: $V_{\Gamma}(x + \gamma) = V_{\Gamma}(x)$, $\forall  x \in \R^d$, 
$\gamma \in \Gamma \simeq \mathbb Z^d$.
\item[\textbf{(ii)}] $U$ is  sub-quadratic: $\d^\alpha U \in L^\infty(\R^d),\ \forall \alpha\in \N^d, \text {such that }|\alpha|\geq 2$. 
\end{itemize}
\end{assumption}
\begin{remark}
Remark that these are the same assumptions as used in \cite{CMS}. In particular they include the cases of isotropic harmonic
potentials $U(x)=|x|^2$, as well as those corresponding to anisotropic ones, like $U(x)=\sum \omega_j^2 x_j^2$. Moreover we can also 
take $U$ to be identically zero, or include a linear component such as $E\cdot x$, $E\in \R$, modeling constant electric fields for example.  
\end{remark}
We proceed by recalling Bloch's famous eigenvalue problem \cite{Bl}. 

\subsection{Bloch's eigenvalue problem} From now on we denote by $\mathfrak C$ the \emph{elementary lattice cell}, \ie the centered
\emph{fundamental domain} of the lattice $\Gamma$, \ie
\begin{equation}\label{eq:Y}
\mathfrak C:= \left\{\gamma \in \R^d: \ \gamma=\sum_{l=1}^d \gamma_l \zeta_l, 
\ \gamma_l\in \left[-\frac{1}{2}, \, \frac{1}{2}\right] \right\},
\end{equation}
whereas the corresponding basic cell of the dual lattice will be denoted by $\mathfrak C^*$. 
In solid state physics $\mathfrak C^*$ is usually called the (first) \emph{Brillouin zone} hence we 
shall also write $\mathfrak B \equiv \mathfrak C^*$. 
Also let us introduce the so-called \emph{Bloch Hamiltonian} (or shifted Hamiltonian) given by
\begin{equation}
\label{bham}
H_{\Gamma}(k): = \frac{1}{2} \, \left(-i\nabla_y + k \right)^2+
V_{\Gamma}\left (y\right), 
\quad k\in \R^d.
\end{equation}
Then Bloch's eigenvalue problem is given by the following spectral cell equation:
\begin{equation}
\label{bloch}
\left \{
\begin{aligned}
H_{\Gamma}(k) \chi_n(y,k) = & \ E_n(k)\chi_n(y,k),\quad \ n\in \N,\, y\in \mathfrak C,\\
\chi_n(y+\gamma,k)=& \ \chi_n(y,k), \quad \quad \quad \quad \mbox{for $\gamma \in \Gamma$}.
\end{aligned}
\right.
\end{equation}
and $E_n(k)\in \R$, $k\in \mathfrak B$, is then called the $n$-th \emph{Bloch eigenvalue} corresponding to the 
potential $V_\Gamma$. We shall now briefly collect some well known facts for this eigenvalue problem, \cf
\cite{Ne, ReSi, Wi}: 
\newpar 
Since $V_\Gamma$ is smooth and periodic, $H_\Gamma(k)$, for every fixed $k\in \mathfrak B$, is self-adjoint on
$H^2(\mathbb T^d)$, $\mathbb T^d = \R^d/\Gamma$, with compact resolvent.  
Hence its spectrum is given by
\begin{equation*}
\sigma (H_\Gamma(k))= \{
E_n(k)\in \R\ ;\ k\in \mathfrak B, \, n\in \N^*\}. 
\end{equation*}
The eigenvalues $E_n(k)$ can then be ordered according to their magnitude and multiplicity, \ie 
$$
E_1(k)\leq\ldots\leq E_n(k)\leq E_{n+1}(k)\leq \dots
$$
Moreover every $E_n(k)$ is periodic w.r.t. $\Gamma^*$ and it holds $E_n(k)=E_n(-k)$. The set 
$$
\{E_n (k)\in \R: E_n(k)\leq E_{n+1}(k), \, k \in \mathfrak B\}
$$ 
is then usually named the $n$th-\emph{energy band} (or Bloch band). 
The associated eigenfunction, the so-called \emph{Bloch waves}, $\chi_n(y,k)$ 
form (for every fixed $k\in\mathfrak B$) a complete orthonormal basis
in $L^2(\mathfrak C)$ and are smooth w.r.t. $y\in \mathfrak C$. For the following we choose the usual normalization  
\begin{equation}
\label{norm}
\left <\chi_n(\cdot, k), \chi_m(\cdot, k) \right>_{L^2(\mathfrak C)}\equiv 
\int_\mathfrak C \overline{\chi_n(y, k)}\chi_m(y, k)dy =
\delta_{n,m},\quad n,\,m\in\N. 
\end{equation} 
Regularity of the $\chi_n$ w.r.t. their dependence on $k\in \mathfrak B$ is more subtle. 
It has been shown \cite{Ne} that for any $n\in \N$ there exists a closed subset
$\mathfrak A \subset \mathfrak B$ such that $E_n(k)$ is analytic. Similarly the eigenfunctions 
$\chi_n(\cdot, k)$ are found to be analytic and quasi-periodic in $k$, 
for all $k \in \mathfrak O:= \mathfrak B \backslash \mathfrak A$. Moreover it holds that 
\begin{equation}
\label{iso}
E_{n-1}(k) < E_n(k) < E_{n+1}(k),\quad \forall \, k \in \mathfrak O.
\end{equation}
If this condition is satisfied for all $k\in \mathfrak B$ then $E_n(k)$ is said to be an \emph{isolated Bloch band} \cite{Te}. 
Finally we remark that the set where one encounters the so-called \emph{band crossings}, is indeed of measure zero, \ie 
$$
\meas \mathfrak A = \meas \, \{ k\in\mathfrak B\ | \ E_n(k)=E_{m}(k), \ n\not = m \}=0.
$$ 
\begin{remark} Note that in the case $d=1$ all band crossings can be
removed through a proper analytic continuation of the bands, \cf
\cite{ReSi}.   
\end{remark}
From the eigenvalue equation \eqref{bloch} we obtain the following useful identities: 
Differentiating the \eqref{bloch} w.r.t. to $k$ (assuming for the moment that everything is sufficiently smooth) 
yields
\begin{equation}
\label{id0}
(\nabla_k H_{\Gamma}(k)- \nabla_k E_n(k))\chi_n+
(H_{\Gamma}(k)- E_n(k))\nabla_k \chi_n=0.
\end{equation}
Hence, taking the scalar product with $\chi_n$, we obtain a formula for $\nabla_k E_n(k)$ by
\begin{equation}
\label{id1}
\begin{aligned}
\left< \chi_n , \,  \nabla_k H_{\Gamma}(k)\chi_n \right>_{L^2(\mathfrak C)}
\equiv & \, \left<  \chi_n , \,  (-i \nabla_y+k) \chi_n  \right>_{L^2(\mathfrak C)} \\
= & \, \nabla_k E_n(k),
\end {aligned}
\end{equation}
since $H_\Gamma $ is self-adjoint. Similarly we obtain the following expression for the 
entries of the Hessian matrix $D^2 E_n(k)$
\begin{equation}
\label{hess}
\begin{aligned}
\partial^2_{k_j k_l} E_n(k) = & \ \delta_{j,l}+  \langle \chi_n , \, ( -i \partial_{y_j} + k_j) \chi_n  \rangle 
 + \langle  \chi_n , \, (-i \partial_{y_l} +k_l) \partial_{k_j}\chi_n\rangle\\
& \ - \langle  \chi_n , \,  (\partial_{k_l}E_n(k)) \partial_{k_j} \chi_n + (\partial_{k_j}E_n(k)) \partial_{k_l} \chi_n\rangle,
\end {aligned}
\end{equation}
where $\delta_{j,l}$ denotes the Kronecker symbol for $j,l=1,\dots,d$. (Below we shall assume $E_n(k)$ to be 
a simple eigenvalue which implies sufficient regularity to justify all differentiations above.)
\newpar

\subsection{Existence of smooth solutions for NLS} As a final preparatory step we state a basic existence and 
uniqueness result for NLS of the form \eqref{snls1} (see also \cite{Caz, SuSu} for a general introduction).

\begin{lemma}\label{lemex}
Let Assumption~\ref{aspot} be satisfied, and let $\psi_I\in
\S(\R^d)$, the Schwartz space. Let $s>d/2$. Then there exists $t=t(\e,h)>0$
and a unique solution $\psi \in C(]-t,t[;H^s(\R^d))$ satisfying \eqref{snls1}. 
Moreover, $x^\alpha\psi \in C(]-t,t[;H^s(\R^d))$ for any $\alpha\in\N^d$, $s\in \N$, and the
following conservation law holds:
\begin{equation}
\label{cons}
\frac{d}{dt}\,{\|\psi(t)\|}_{L^2} =0\, .
\end{equation}
\end{lemma}
\begin{proof}
See the proof of Lemma 4.3 in \cite{CMS}.
\end{proof}

\begin{remark}
In general, one cannot expect global-in-time existence for solutions to NLS. For example, if
$\kappa$ is negative and if $\sigma > 2/d$, finite time blow-up may occur, see \eg \, \cite{Car, SuSu} (see 
also \cite{Caz} for the case $\sigma = 2/d$). 
\end{remark}


\section{Multiple scales expansion}\label{sexp}

We now establish the asymptotic behavior of $\psi(t)$, solution to \ref{snls1}, for $0< \e \ll 1$, by means of a 
multiple scales expansion. In the following $h >0$ is kept fixed, though we shall not simply put it equal to $1$, 
since we want to keep track of its appearance in order to compare our results with the semi-classical situation of \cite{CMS}, 
\cf Remark \ref{screm} below. 
Similar to \cite{AlPi}, we shall first consider the easier situation where no large drifts appear and 
then include them in a second step, using a more general asymptotic expansion method.

\subsection{Homogenization without drift} 
In this sub-section we seek an asymptotic expansion for solutions to \eqref{snls} in the following form:  
\begin{equation}
\label{multi}
\left \{
\begin{aligned}
& \, \psi(t,x)  =   \, u^\e \left(t,x,\frac{x}{\e}\right)\exp \left( i\left(\frac{k_0 \cdot x}{\e} + \frac{\beta t}{\e^2}\right) \right),\\
& \, u^\e (t,x,y) \sim  \ \sum_{j=0}^\infty \e^j u_j(t,x,y), 
\end{aligned}
\right.
\end{equation}
where $k_0 \in \R^d$ is induced by the initial condition and $\beta \in \R$ is some arbitrary constant to 
be determined below. The precise meaning of the symbol ``$\sim$'' in terms of an asymptotic series 
will be discussed in Section \ref{sstab} below. Moreover we impose that $u^\e (t,x,y)\in \C$ satisfies
\begin{equation*}
u^\e(\cdot,\cdot,y + \gamma) = u^\e(\cdot,\cdot,y), \quad \forall \, y
\in \R^d,  
\, \gamma \in \Gamma,
\end{equation*}
in order to capture precisely those oscillations which are introduced via $V_\Gamma$.
\begin{remark}
This particular form of a multiple-scale ansatz is suggested by the linear results given in \cite{ACPSV, AlPi}.
Indeed one could have started with a more general ansatz, imposing appropriate periodicity or quasi-periodicity 
assumptions. It then turns out that one ends up with again the same form as given in \eqref{multi}. Also, the ansatz 
\eqref{multi} might not be so surprising when compared to the \emph{two-scale WKB-approach} used in 
\cite{BLP,CMS, GRT} (see also the Remark \ref{screm} below).
\end{remark}
As usual in asymptotic expansion methods we have to henceforth assume that the initial condition $\psi_I$ is 
compatible with \eqref{multi}.
\begin{assumption}[Well-prepared initial data I]
\label{assi} 
The initial data $\psi_I$ is assumed to be of the following form:
\begin{equation}
\psi_I(x)= u^\e_I\left(x,\frac{x}{\e}\right)e^{i k_0\cdot x/\e} , \quad \mbox{for some given $k_0\in \R$,}
\end{equation}
where $u^\e_I(x,y)$ is $\Gamma$-periodic w.r.t. to $y$ and $u_I^\e\in \mathcal S(\R^{2d})$\footnote{That is, $u^\e_I$ is smooth 
and rapidly decaying w.r.t. $x$ and smooth w.r.t. $y$.}.
\end{assumption}
Assuming for the moment that $u^\e$ is sufficiently smooth we (formally) plug the ansatz \eqref{multi} into \eqref{snls1} and compare 
equal powers in $\e$. This yields
\begin{equation}
\label{epeq}
\frac{1}{\e^2} L_0 u^\e + \frac{1}{\e} L_1 u^\e + L_2 u^\e + \kappa |u^\e|^{2\sigma} u^\e= 0,
\end{equation}
where the linear differential operators $L_0$ and $L_1$ are defined by 
\begin{equation}\label{L0}
\begin{aligned}
\begin{aligned}
L_0 u^\e & :=  h \beta + h^2 H_{\Gamma}(k_0),\\
L_1u^\e & := - h^2 \nabla_x \cdot \nabla_y u^\e + i h^2 k_0\cdot \nabla_x u^\e,
\end{aligned}
\end{aligned}
\end{equation}
with $H_\Gamma(k)$ being the Bloch Hamiltonian as given in \eqref{bham}. We also define
\begin{equation}
\label{L2}
L_2 u^\e := -i h \, \partial_t u^\e - \frac{h^2}{2} \, \Delta_x u^\e + U(x) u^\e.
\end{equation}
Since $u^\e \sim \sum \e^j u_j$ we shall in the following expand equation \eqref{epeq} in powers of $\e$ and derive conditions 
on $u_j$, such that all resulting terms are zero up to sufficient high orders in $\e$. 
\newpar
Setting the leading order term, \ie the term of order $\O(\e^{-2})$, equal to zero gives 
\begin{equation}
\label{0eq}
H_\Gamma(k_0)u_0+ \frac{\beta}{h} \, u_0=0,  
\end{equation}
from which we readily see that we need to choose 
\begin{equation}
\beta = -h E_n(k_0). 
\end{equation}
Now, if we assume that $E_n(k_0)$ is indeed a \emph{simple} Bloch-eigenvalue then \eqref{0eq} implies that $u_0$ 
can be decomposed as
\begin{equation}
\label{u0}
u_0(t,x,y) = f_0(t,x)\chi_n(y,k_0),\quad \forall \, t\in \R,x\in \R^d,
\end{equation} 
with some yet undetermined function $f(t,x)\in \C$. This now leads directly to the following important assumption
\begin{assumption}[Well-prepared initial data II]
\label{assu}
Initially, the leading order ``amplitude'' $u_0$ is assumed to be concentrated in a single 
Bloch band $E_n(k_0)$ corresponding to a simple eigenvalue of $H_\Gamma(k_0)$, \ie  
\begin{equation}
\label{u}
u_0(0,x,y)\equiv f_0(0,x) \chi_n(y,k_0),
\end{equation}
where $f_0(0,\cdot)\equiv f_I(\cdot)\in \mathcal S(\R^d; \C)$ is some given initial data.
\end{assumption}
An important consequence of $E_n(k_0)$ being simple is that that in this case it is known to be infinitely differentiable in a vicinity of $k_0$.
\newpar
We have seen that in leading order $\psi(t)$ can be written as
\begin{equation}
\label{princ}
\psi(t,x)\sim f_0(t,x) \chi_n\left(\frac{x}{\e},k_0\right)e^{i k_0\cdot x/\e}e^{-ih  E_n(k_0)t/\e^2} +\O(\e),
\end{equation}
where the $f_0$ is yet to be determined. To this end we proceed with our asymptotic expansion by 
setting terms of order $\O(\e^{-1})$ equal to zero. This yields  
\begin{equation}
\label{eq}
(H_\Gamma(k_0)- E_n(k_0)) u_1= \nabla_x \cdot \nabla_y u_0 -i k_0\cdot \nabla_x u_0,
\end{equation}
which by inserting \eqref{u0} can be rewritten as
\begin{equation}
\begin{aligned}
\label{1eq}
(H_\Gamma(k_0)- E_n(k_0)) u_1= & \, -i\nabla_x f_0 \cdot (-i\nabla_y \chi_n + k_0 \chi_n),\\
= & \, -i\nabla_x f_0 \cdot \nabla_k H_{\Gamma}(k_0) \chi_n,
\end{aligned}
\end{equation}
where the second equality follows from the definition of $H_\Gamma(k)$ \eqref{bham}. 
It remains to be asked if this equation is solvable for $u_1$. By Fredholm's alternative the necessary and 
sufficient condition to do so is that the right hand side is orthogonal (in $L^2(\mathfrak C)$) to $\chi_n(y,k_0)$. Hence we have to impose that 
\begin{equation}
\begin{aligned}
0= & \, -i\nabla_x f_0 \cdot\left< \chi_n , \,  \nabla_k H_{\Gamma}(k_0)\chi_n \right>_{L^2(\mathfrak C)}\\
= & \, -i\nabla_x f_0 \cdot \nabla_k E_n(k_0),
\end{aligned}
\end{equation}
where for the second equality we used the identity \eqref{id1}. Thus we are lead to the restriction that $k_0$ has to be 
a critical point of $E_n(k)$, \ie  
\begin{equation}
\label{vcon}
\nabla_k E_n(k_0)=0.
\end{equation} 
This situation is analogous to the one discussed in the first part of \cite{AlPi}, though the arguments given there are different. 
Assuming that \eqref{vcon} indeed holds true, we get from \eqref{1eq}, together with \eqref{id0}, that the 
order $\O(\e)$-\emph{corrector} $u_1$ in general can be written in the following form 
\begin{equation}
\label{u1}
u_1(t,x,y) =  -i \nabla_x f_0(t,x) \cdot \nabla_k \chi_n(y,k_0) + f_1(t,x) \chi_n(y,k_0),
\end{equation}
for any given function $f_1$. 
Remark that we can not choose $u_1(0,x,y)$ completely arbitrary, once $u_0(0,x,y)$ is fixed, \ie 
the initial data $u_I^\e$ given in Assumption \ref{assi} (formally) has to be of the following form
\begin{equation}
\label{leou}
u^\e_I(x,y)\sim  (f_0(0,x)+\e f_{1}(0,x)) \chi_n(y,k_0) -i \e \nabla_x f_0(0,x) \cdot \nabla_k \chi_n(y,k_0) +  \O(\e^2),
\end{equation}
in order to be consistent with our asymptotic description. In the following we shall choose $f_1(0,x)\equiv 0$ for 
simplicity.
\newpar
Proceeding with our $\e$-expansion of \eqref{epeq}, we next consider terms of order $\O(1)$ to obtain the following equation:
\begin{equation}
\label{eq2}
L_0 u_2 + L_1 u_1+ L_2 u_0 + \kappa |u_0|^{2\sigma} u_0 = 0.
\end{equation}
Again, by Fredholm's alternative, it can be solved for $u_2$, iff 
\begin{equation}
\int_{\mathfrak C} \overline {\chi_n(y,k_0)} \left(L_1 u_1+ L_2 u_0 + \kappa |u_0|^{2\sigma} u_0 \right) dy = 0.
\end{equation}
Plugging into this identity the precise forms of $u_0$ and $u_1$, respectively defined in \eqref{u0} and \eqref{u1}, and using 
formula \eqref{hess}, given the fact that $\nabla_k E_n(k_0)=0$, we obtain after some lengthy but straightforward calculations 
the following solvability 
condition:
\begin{equation}
\label{effnls}
\left \{
\begin{aligned}
i h \partial _t f_0 = & \  -\frac{h^2}{2} \, \diverg_x(M^*\nabla_x) f_0 + U(x)f_0 + 
\kappa^* \, |f_0|^{2\sigma} f_0,\\
f_0 \big |_{t=0}   = &  \ f_I(x).
\end{aligned}
\right.
\end{equation}
This is nothing but the \emph{homogenized NLS}, or the \emph{effective mass NLS}, where the so-called \emph{effective mass tensor} $M^*\in \R^{d\times d}$ 
is defined by
\begin{equation}
\label{m}
M_{j,l}^*:= \partial^2_{k_j,k_l}E_n(k_0),\quad j,l=1,\dots,d.
\end{equation}
Moreover the \emph{effective coupling constant} $\kappa^*\in \R$ within the $n$-th Bloch band is defined by 
\begin{equation}
\label{k}
\kappa^*(k_0) :=  \kappa \int_{\mathfrak C}\left|\chi_n\(y,k_0\)\right|^{2\sigma +2} \, dy .   
\end{equation}
The effective NLS \eqref{effnls} describes the dispersive dynamics, as $\e \rightarrow 0$, of \eqref{snls1} for long macroscopic time-scales.  
However, it should not be confused with the so-called effective Hamiltonian as determined in \cite{Ne, Te}. 
Note that in general $M^*$ is \emph{neither positive nor definite}, thus equation \eqref{effnls} in general also includes the class 
of so-called \emph{non-elliptic NLS} \cite{GhSa, KPV, SuSu}.
\begin{remark} 
The formulas \eqref{effnls}-\eqref{k} can be checked to be exactly the same as in the physics literature \cite{PBZBM, StZh} 
and moreover simplify to the ones given in \cite{AlPi, PoRi} in the linear case. If $M^*$ is scalar its inverse $m^*=1/M^*$ is 
called the \emph{effective mass}.
\end{remark}
In the next subsection we shall show how to get rid of the additional assumption \eqref{vcon} that $k_0$ is a critical point of $E_n(k)$.

\subsection{General situation including drifts} In order to generalize the expansion to situations where $\nabla_k E_n(k)\not =0$ we 
have to modify our multiple-scales ansatz. It turns out that instead of \eqref{multi} we need to consider
\begin{equation}
\label{multig}
\begin{aligned}
\psi(t,x)  \sim  u^\e \left(t,\tilde x,\frac{x}{\e}\right)\exp \left( i\left(\frac{k_0 \cdot x}{\e} - \frac{h E_n(k_0)t}{\e^2}\right) \right),
\end{aligned}
\end{equation}
where $u^\e \sim \sum \e^j u_j$ and the new spatial coordinate $\tilde x$ is given by
\begin{equation}
\label{drift}
\tilde x := x - \frac{h}{\e}\, \omega(k_0) t,\quad \mbox{with }\omega(k_0):=\nabla_k E_n(k_0).
\end{equation}
Thus $\tilde x$ comprises a \emph{macroscopically large drift} with a drift-velocity proportional to $\nabla_k E_n(k_0)$. 
Note that the fast scale $x/\e$ remains unchanged, hence in situations where $\nabla_k E_n(k) =0$ we are clearly back to our old situation. 
Similarly as before we plug \eqref{multig} into \eqref{snls1} which yields
\begin{equation}
\label{epeqg}
\frac{1}{\e^2} L_0 u^\e + \frac{1}{\e} \tilde L_1 u^\e + L_2 u^\e + \kappa |u^\e|^{2\sigma} u^\e= 0,
\end{equation}
where the linear differential operator $L_0$, $L_2$ are defined as in \eqref{L0}, \eqref{L2}, respectively, 
but with $x$ replaced by $\tilde x$. However instead of $L_1$ we have  
\begin{equation}
\begin{aligned}
\tilde L_1u^\e:= - h^2 \nabla_{\tilde x} \cdot \nabla_y u^\e + i h^2 (k_0+\omega(k_0))\cdot \nabla_{\tilde x} u^\e,
\end{aligned}
\end{equation}
Then, by exactly the same arguments as above we obtain that the leading order amplitude is given by 
\begin{equation}
\label{u0g}
u_0(t,\tilde x,y) = f_0(t,\tilde x)\chi_n(y,k_0),\quad \forall \, t\in \R,x\in \R^d.
\end{equation} 
However, setting next the $\O(\e^{-1})$-term equal to zero yields, instead of \eqref{1eq},
\begin{equation}
\label{1eqg}
(H_\Gamma(k_0)- E_n(k_0)) u_1=  \, -i\nabla_x f_0 \cdot (\nabla_k H_{\Gamma}(k_0) \chi_n - \omega(k_0) \chi_n).
\end{equation}
In this case, the solvability condition requires
\begin{equation}
\label{solg}
\begin{aligned}
0= & \, \left< \chi_n , \,  (\nabla_k H_{\Gamma}(k_0)- \omega(k_0))\chi_n \right>_{L^2(\mathfrak C)}\\
=& \, \nabla_k E_n(k_0)- \omega(k_0),
\end{aligned}
\end{equation}
where we have used the normalization $\left< \chi_n ,\chi_n \right>_{L^2(\mathfrak C)} =1$. But 
\eqref{solg} of course is identically fulfilled by definition of $\omega(k_0):=\nabla_k E_n(k_0)$. 
Thus, by identity \eqref{id0}, we formally obtain the same $\O(\e)$-corrector $u_1$ 
\begin{equation}
\label{u1g}
u_1(t,\tilde x,y) =  -i \nabla_x f_0(t,\tilde x) \cdot \nabla_k \chi_n(y,k_0) + f_1(t,\tilde x) \chi_n(y,k_0),
\end{equation}
for any given function $f_1$. (Like above we set $f_1$ to be identically zero at $t=0$, for simplicity.) 
Thus by transforming the $x$-coordinate 
into $\tilde x$ by \eqref{drift} we can now proceed with our asymptotic expansion, 
having gained the additional freedom to include the case $\nabla_k E_n(k_0)\not =0$. 
The equation of order $\O(1)$ now gives
\begin{equation}
\label{eq2g} 
L_0 u_2 +\tilde L_1 u_1+ L_2 u_0 + \kappa |u_0|^{2\sigma} u_0 =0.
\end{equation}
It is clear now that as before, the corresponding solvability condition \ie 
\begin{equation}
\label{eq2gsol}
\int_{\mathfrak C} \overline {\chi_n(y,k_0)} \left(\tilde L_1 u_1+ L_2 u_0 + \kappa |u_0|^{2\sigma} u_0 \right) dy = 0,
\end{equation}
yields a homogenized NLS equation of the same form as in \eqref{effnls}, namely
\begin{equation}
\label{effnlsg}
i h \partial _t f_0 = \,  -\frac{h^2}{2} \, \diverg(M^*\nabla) f_0 + U(x)f_0 + 
\kappa^* \, |f_0|^{2\sigma} f_0.
\end{equation}
One can easily check that even though $\nabla_k E_n(k_0)\not =0$ in this case, all additional terms appearing in \eqref{eq2gsol}, cancel 
out identically, hence equation \eqref{effnlsg} remains.
\begin{remark} In the physics literature \cite{PBZBM, StZh} the variable-transformation $x\rightarrow \tilde x:= x - h\omega(k_0) t/\e$ 
is sometimes reverted, leading to an additional convective term on the left hand side of \eqref{effnlsg}. This however 
can be considered only as a formal statement since consequently the large factor $\e^{-1}$ would appear then in the 
homogenized NLS, a somewhat inconsistent formalism. 
\end{remark}
To prove the existence of smooth solutions for the homogenized NLS we shall impose the following ellipticity-assumption:
\begin{assumption}[Ellipticity]
\label{assm} 
We assume that at $k_0 \in \mathfrak B$ it holds: 
\begin{equation}\label{ell}
\xi^T M^* \xi \equiv \sum_{k,l=1}^d \partial^2_{k_j k_l} E_n(k_0)\xi_j \xi_l \geq C |\xi|^2,\quad \mbox{for $\xi \in \R^d$, $C>0$.}
\end{equation}
\end{assumption}
Clearly, condition \eqref{ell} is valid if $k_0\in \mathfrak B$ is indeed a local minimum of $E_n(k)$. It may very well be possible to relax 
the above assumption, \cf Remark \ref{remell} below. Here we mainly introduced it for definiteness, since under the 
condition \eqref{ell} it is then straightforward to prove that the effective NLS \eqref{effnls} (or equivalently \eqref{effnlsg}) 
has a smooth solution, at least locally-in-time.
\begin{lemma}\label{lemex1}
Let the Assumptions~\ref{aspot} and \ref{assm} be satisfied, and let $f_I\in
\S(\R^d)$. Then there exists $\tau=\tau(h)>0$ and a unique solution $f_0 \in C(]-\tau,\tau[;H^s(\R^d))$, $s>d/2$, satisfying 
\eqref{effnls}, or equivalently \eqref{effnlsg}. Moreover, $x^\alpha f_0 \in C(]-\tau,\tau[;H^s(\R^d))$ for any 
$\alpha\in\N^d$, $s\in \N$.
\end{lemma}
\begin{proof} By Assumption \eqref{assm} we have that $-\diverg (M^* \nabla)$ is uniformly elliptic and since $U$ is sub-quadratic 
the operator $-\diverg (M^* \nabla)+U$ is therefore well known to be essentially self-adjoint on $C_0^\infty(\R^d)$, \cf \cite{ReSi}. The existence of 
a smooth solution $f_0(t,\cdot)\in H^s(\R^d)$, $s>d/2$, therefore follows by the same arguments as it does for the standard NLS \cite{Caz}. The 
higher order regularity is then also proved similarly to, \eg, \cite{Caz, HNT} (see also the proof of Lemma 4.3 in \cite{CMS} for the main 
strategy). 
\end{proof}
\begin{remark}\label{remell}
Indeed one can expect similar existence results to be true under much weaker conditions as given by \eqref{ell}, \cf \cite{KPV}. It is beyond the 
scope of this work though to study the weakest possible assumptions needed (including for example also degenerate cases) but rather 
we refer the reader to \cite{SuSu} and 
the references given therein. Here we only remark that in the case where $M^*$ is \emph{diagonal}, the existence of smooth local-in-time solutions 
is known even in the non-elliptic case \cite{GhSa}. 
\end{remark}
Thus, at least for $t\in ]-\tau, \tau[$, we have that $u_0$, and hence also $u_1$, is smooth w.r.t. $x,y$ and moreover 
in $H^s(\R^d)$ w.r.t. to $x$ for any $s\in \N$. 
It is clear that in general we can not expect smooth global-in-time solutions of the effective NLS \eqref{effnls}. 
A situation though were one indeed has globally smooth solutions, \ie $\tau = \infty$, is furnished by condition \eqref{ell} 
together with assumption \ref{aspot} and imposing that in addition $\kappa >0$. 
\begin{remark} \label{screm}
Let us briefly compare the obtained leading order asymptotic description of $\psi(t)$ with the one 
derived in \cite{CMS} for the weakly nonlinear semi-classical scaling: Indeed if we formally set $h=\e$ in \eqref{princ}, \eqref{u0g}, 
we obtain
\begin{equation}
\psi(t,x)\sim f_{0}(t,x-\omega(k_0)t) \chi_l\left(\frac{x}{\e},k_0\right)e^{i (k_0\cdot x- E_n(k_0)t)/\e} +\O(\e),
\end{equation}
which is exactly of the same form as the two-scale WKB-ansatz used in \cite{CMS} (see also \cite{BLP,GRT} for the linear case). 
In this case the highly oscillatory WKB-phase is simply given by $\phi(t,x)= k_0\cdot x- E_n(k_0)t$. Note that this $\phi$ solves
the $n$-th band \emph{semi-classical Hamilton-Jacobi equation} with vanishing external field and plane wave initial data, \ie
\begin{equation}
\label{hj}
\left \{
\begin{aligned}
& \partial_t \phi + E_n(\nabla_x \phi) = 0 ,\\ 
& \phi \big|_{t=0} = k_0\cdot x.
\end{aligned}
\right.
\end{equation}
Moreover with this choice of $\phi$ (and since $U$ vanishes) one easily checks that the transport equation for 
the leading order WKB-amplitude, as derived in \cite{CMS}, simplifies to
\begin{equation}
\label{trans}
\left \{
\begin{aligned}
& \partial_t f_0 + \omega(k_0)\cdot \nabla_x f_0 = 0,\\ 
& f_0\big |_{t=0}= f_I(x).
\end{aligned}
\right.
\end{equation}
Clearly, the solution of \eqref{trans} is then simply given by 
\begin{equation}
f_0(t,x) = f_I(t,x-\omega(k_0)t),
\end{equation}
and hence consistent with our approach.  
\end{remark}
\subsection{Higher order expansions} We can henceforth proceed with our $\e$-expansion of equation \eqref{epeq}. Denote the projector onto the 
$n$th-Bloch band corresponding to a simple eigenvalue $E_n(k)$ by 
$$\mathbb P_n(k):=|\chi_n(y,k)\rangle \langle \chi_n(y,k)|,$$ 
(using the convenient Dirac notation) and moreover define 
$$\mathbb Q_n(k):=\id - \mathbb P_n(k).$$ 
This operator is smooth in 
a vicinity of $k_0$ and hence, by elliptic inversion, a partial inverse for $L_0\equiv L_0(k_0)$ can be defined on its range, \ie 
$L_0^{-1}\mathbb Q(k_0)$ is well-defined, and smooth. Coming back to equation \eqref{eq2g} we can decompose $u_2$ as 
\begin{equation}
\label{u2}
u_2(t,x,y)= f_2(t,x)\chi_n\(y,k_0)\)+ u_2^\perp(t,x,y),
\end{equation}
where the function $f_2$ is yet unknown and $u_2^\perp$ is such that
\begin{equation*}
\mathbb P_n(k_0) u_2^\perp(t,x,\cdot)  = \langle \chi_n(\cdot,k_0), u_2^\perp(t,x,\cdot)\rangle_{L^2(\mathfrak C)} = 0, \quad  
\forall \, (t,x)\in ]-\tau,\tau[\times \R^d.
\end{equation*}
Now, $u^\perp_2$ is determined by \eqref{eq2g} via 
\begin{equation}\label{u_1per}
u^\perp_2 = -L_0^{-1}\mathbb Q_n(k_0) \(\tilde L_1 u_1+ L_2 u_0 + \kappa |u_0|^{2\sigma} u_0)\),
\end{equation}
which implies $u^\perp_2\in C(]-\tau,\tau[;H^s(\R^d))$, since $u_0$ and $u_1$ respectively are, by
Lemma~\ref{lemex1}. As before, equation \eqref{u_1per} henceforth induces a particular form of the $\O(\e^2)$-corrector 
in the initial amplitude $u^\e_I\sim \sum \e^j u_j$. The next higher order in $\e$ leads us to the 
following \emph{linear} problem (after a Taylor-expansion of the nonlinearity around $u_0$):
\begin{equation}
\label{eq3}
L_0 u_3 + \tilde L_1 u_2 + L_2 u_1 + \kappa \left((2\sigma +1)|u_0|^{2\sigma} u_1 + 2\sigma |u_0|^{2\sigma -2} u_0^2\bar u_1\right) = 0.
\end{equation}
The corresponding solvability condition then determines $f_1(t,x)\in \C$, \ie the amplitude correpsonding 
to the \emph{polarized part} of the first order amplitude 
$u_1(t,x,y)$ given in \eqref{u1}. This then leads to a homogenized \emph{linear} Schr\"odinger-type equation 
for $f_1(t,x)$, where we have the freedom to choose $f_1(0,x)=0$. 
\newpar
By this procedure, all higher order terms $u_j(t,x,y)$, $j\geq 1$, of
the asymptotic solution \eqref{multig} can be obtained and it is now clear that we can always choose $g_j(0,x)$, 
\ie the polarized part of $u_j(0,x,y)$, to be identically zero. In the \emph{globally periodic case}, \ie $U(x)=0$ and $\kappa=0$, 
the non-vanishing higher order terms $u_j^\perp(t,x,y)$, $j\geq 1$, are found to be combinations of 
higher order derivatives w.r.t. $k$ and $x$ respectively. 
of $\chi_n$ and $f_0$, \cf \cite{BLP, CoVa}. Although this is no longer true in our case we still have that 
$u_j\in C(]-\tau,\tau[;H^s(\R^d))$ for all $j\geq 1$. 
At each step though, an additional condition is imposed (recursively) for the initial data $\psi_I$. 
This can be seen analogous to the situation encountered in \cite{CMS} and can be understood in the framework of 
so-called \emph{super-adiabatic subspaces} as constructed in \cite{PST}. 

\begin{remark} 
Note that the above given construction can be extended to the case where $E_n(k)$ is an \emph{$m$-fold degenerate}
family of eigenvalues, \ie  
$$
E_n(k_0) = E_*(k_0),\quad \forall \, n\in I\subset \N,\, |I|=m,
$$ 
under the additional assumption that there exists a smooth orthonormal basis $\chi_l(y,k_0)$, $l=1,\dots,m$, 
of $\ran \mathbb P_I(k)$, where $\mathbb P_I(k_0)$ denotes the spectral projector corresponding to $E_*(k)$. In this case 
the leading order asymptotic description would be
\begin{equation}
\psi(t,x)\sim \sum_{l=1}^m f_{0,l}(t,\tilde x) \chi_l\left(\frac{x}{\e},k_0\right)e^{i k_0\cdot x/\e}e^{-ih  E_n(k_0)t/\e^2} +\O(\e),
\end{equation}
In this case however we would be forced to consider a \emph{coupled system of homogenized equations}. The corresponding analysis is 
then analogous to the given one but requires rather tedious computations, a situation which we wanted to avoid for simplicity. 
For an extensive study of such situations in the linear case we refer to the last section of \cite{AlPi}. 
\end{remark}


\section{Nonlinear stability of the asymptotic solution}\label{sstab} 

To prove that the above given multiple-scale expansion indeed yields a good approximation of the
exact solution $\psi(t)$ for $\e \ll 1$, a nonlinear stability result is needed. Note that due to the scaling 
of \eqref{snls1} we can not hope for any uniform (w.r.t. $\e$) bound in, say, $H^s(\R^d)$ for $\psi(t)$. 
On the other hand the uniform $L^2$ estimate \eqref{cons} is clearly not sufficient to pass to the limit in the nonlinearity. 
This motivates the introduction of the following $\e$-scaled spaces:
\begin{definition} For $s\in \N$ let
\begin{equation*}
Y^s_\e :=\left\{ f^\e \in L^2(\R^d)\ ;\ \sup_{0<\e\leq 1}{\|f^\e\|}_{Y^s_\e}  <+\infty\right\},
\end{equation*}
where 
\begin{equation*}
{\|f^\e\|}_{Y^s_\e}:=
\sum_{|\alpha|+|\beta|\leq s}\left\|
(\e x)^\alpha (\e \d)^\beta f^\e\right\|_{L^2(\R^d)}.
\end{equation*}
\end{definition}
\begin{remark}
Similar spaces, but without the extra weight $\e^\alpha$ have been used in the semi-classical study given in \cite{CMS}. 
Both variants can be seen as an extension of the $H^s_\e$-spaces, defined by 
\begin{equation}
{\|f^\e\|}_{H^s_\e}:=
\sum_{|\beta|\leq s}\left\|(\e \d)^\beta f^\e\right\|_{L^2(\R^d)}.
\end{equation}
which in the context of geometrical optics expansion have been first introduced in \cite{Gu} (see also \cite{Ra} and the 
references given therein). In our case the additional factor $(\e x)^\alpha$ is needed because we want to include sub-quadratic 
potentials $U(x)$ (and due to our scaling we can not work in the $X^s_\e$-spaces introduced in \cite{CMS}). 
If we would allow $U(x)$ to grow only only sub-linearly we could as well work in $H^s_\e$. 
\end{remark}
\newpar
\textbf{Notation.} Let $(\alpha^\e)_{0<\e\leq 1}$ and $(\beta^\e)_{0<\e\leq 1}$ be  
two families of positive numbers. From now on we shall write
\begin{equation*}
\alpha^\e \lesssim \beta^\e,
\end{equation*}
if there exists a $C>0$, independent of $\e \in ]0,1]$ (but possibly dependent on other parameters), such that 
\begin{equation*}
\alpha^\e \leq C \beta^\e,\quad \text{for all } \e \in ]0,1].
\end{equation*} 
Since for the following results the value of $h$, appearing in \eqref{snls1}, is indeed irrelevant we shall set $h\equiv 1$ throughout this section. 
Moreover, we shall no longer distinguish between the usual spatial coordinate $x$ and its shifted value $\tilde x$, since our results apply in both situations. 
\newpar
Next, let us specify precisely the class of well-prepared initial data which we need to consider.
\begin{assumption}[Well-prepared initial data III]
\label{assw}
The initial data $\psi_I^\e$ satisfies Assumptions~\ref{assi} and
\ref{assu}, such that for some $K\in \N$, it holds
\begin{equation}\label{eq:K}
u_I^\e(x)= \sum_{j=0}^K \e^{j} u_j(x,y)\Big|_{y=x/\e}
+\O\(\e^{K+1}\), \quad \mbox{in $Y^s_\e$ for any $s\in \N$.}
\end{equation}
Moreover, with $u_0$, $u_1$ given by \eqref{u}, \eqref{leou} respectively, and with $F(z):=|z|^{2\sigma} z$, the
functions $u_j$, $j\geq 2$, are recursively given by 
\begin{equation*}
u_{j} = -  L_0^{-1}\mathbb Q(k_0) \( \tilde L_1 u_{j-1}  +L_2
u_{j-2} + \kappa \frac{d^{j-2}}{ds^{j-2}} F \Big(u_0 + \sum_{l=1}^{j-2} s^l u_l\Big)\Big|_{s=0}\).
\end{equation*}
\end{assumption}
After what we have encountered in the construction of higher order asymptotic solutions, this assumption should not come as a surprise. 
In the linear case the Assumption \ref{assw} is needed if one aims for refined asymptotic estimates. As we shall 
see, the inclusion of higher order asymptotics in our case is needed to control the nonlinear term in the proof of the 
stability result. To this end we need the following existence result for well prepared initial data:
\begin{lemma}\label{lem:Borel}
There exists $\psi_I \in \mathcal S(\R^d)$ such that Assumption~\ref{assw} holds true for any $K\in \N$.  
\end{lemma}
\begin{proof} The proof follows from Borel's theorem, \cf Theorem 4.2 in \cite{Ra}.
\end{proof}
For the following, define, the $N$-th order asymptotic solution by 
\begin{equation}\label{eq:v}
v^\e_N(t, x) := \( \sum_{j=0}^N \e^j
u_j\(t,x,\frac{x}{\e}\)\)e^{i k_0\cdot x/\e}e^{-ih  E_n(k_0)t/\e^2},   
\end{equation}
and moreover let 
\begin{equation}
\label{ham}
H^\e: =  - \frac{1}{2} \Delta +
\frac{1}{\e^2} V_{\Gamma}\left(\frac{x}{\e}\right) + U(x)
\end{equation}
denote the linear part of the Hamiltonian operator. (Remark that the scaling of \eqref{ham} is different from the standard semi-classical 
one as used in \cite{CMS, PST}.) In the foregoing section we obtained the following preliminary result:
\begin{proposition}\label{prop}
Let $\psi_I$ satisfy Assumption~\ref{assw} for any $K\in \N$ and $\tau >0$ be the existence-time of smooth solutions 
to \eqref{effnlsg}. Then for any $N\in \N$, $v_N^\e(t)$ solves
\begin{equation}
\label{approx}
\left \{
\begin{aligned}
i \partial _t v_N^\e- H^\e v^\e_N  = & \ \kappa |v_N^\e|^{2\sigma} v_N^\e +\e^{N}r_N^\e,\\
v_N^\e \big |_{t=0}   = &  \  \psi_I + \e^{N+1}\eta_{N+1}^\e,
\end{aligned}
\right.
\end{equation}
where $r_N^\e \in C(]-\tau,\tau[;\, H^s(\R^d))$, $\eta_{N+1}^\e \in \mathcal S(\R^d)$
are such that $r_N^\e \in L^\infty_{\rm loc}(]-\tau ,\tau[; Y^s_\e)$ and 
${\| \, \eta_{N+1}^\e \, \|}_{Y^s_\e}=\O(1)$ for any $s\in \N$. 
\end{proposition}
The main result we shall prove is then given by the following theorem:
\begin{theorem}
\label{theo:stab}
Let $\psi_I$ satisfy Assumption~\ref{assw} for any $K\in \N$, $\tau >0$ be the existence-time 
of smooth solutions to \eqref{effnlsg}, and 
$v_N^\e$ given by \eqref{eq:v}. Then for any $\tau_0 < \tau$, 
there exists $\e_0>0$ such that for $0<\e\leq \e_0$, the solution $\psi(t)$ to \eqref{snls1} is defined on the 
time-interval $[-\tau_0,\tau_0 ]$ and moreover it holds
\begin{equation}\label{eq:O}
\sup_{t\in [-\tau_0, \tau_0]}{\left\| \psi(t) - v_N^\e(t)\right\|}_{Y^s_\e} 
= \O\(\e^{N+1}\),
\end{equation} 
for any $N\in \N$ and $s\in \N$.
\end{theorem}
The above given results shows that if $f(t)$ does not blow up in finite time, then neither does $\psi(t)$, at least for $\e$ 
sufficiently small. Further notice that if $\tau = \infty$, then the above given estimate \eqref{eq:O} holds for any 
bounded time-interval $[\tau_0, \tau_0]\in \R_t$, 
in contrast to the (nonlinear) semi-classical situation \cite{CMS} where the appearance of \emph{caustics} 
usually causes the WKB-approach to break down in finite time.  
Note that in this result, Assumption ~\ref{assw} on the initial data $\psi_I$ has to be valid for any $K\in \N$. We 
shall show in Proposition \ref{prop:finale} below how to relax this restriction. 
\begin{proof} The proof is similar to the one given in \cite{CMS}. Due to the different scaling of our equation, we shall 
present it here in more detail though, which moreover should benefit the reader. 
Define the difference between the exact and the asymptotic solution as 
$$
w^\e_N (t,x):= \psi(t,x) - v^\e_N(t,x).
$$ 
Then, from \eqref{snls1} and \eqref{approx}, $w_N$ solves
\begin{equation}\label{eq:w}
\left\{
\begin{aligned}
i\d_t w^\e_N &= H^\e w^\e_N  + \kappa\(|\psi|^{2\sigma}\psi - |v_N|^{2\sigma}v_N\)- \e^{N}r^\e_N, \\
w^\e_N\big|_{t=0} &=  \e^{N+1}\eta_{N+1}^\e.
\end{aligned}
\right.
\end{equation}
By Lemma \ref{lemex1} and the well known Gagliardo-Nirenberg inequality, we have that 
$v_N^\e$ is uniformly bounded in $L^\infty([-\tau_0,\tau_0]\times\R^d)$. We shall prove now that $w_N^\e$ is also bounded
in $L^\infty([-\tau_0,\tau_0]\times\R^d)$, by using a continuity argument which shows that $w_N\e$ is actually small 
in that space, for $N$ sufficiently large. To this end we first note that the following important relation holds
\begin{equation}
{\|f^\e\|}_{H^s}= \e^{-d/2}{\|f^\e\|}_{H^s_\e} \lesssim \e^{-d/2}{\|f^\e\|}_{Y^s_\e},
\end{equation}
where the scaling factor $\e^{-d/2}$ can be easily seen by Fourier transformation. This then directly 
leads us to an $\e$-scaled Gagliardo-Nirenberg type inequality, \ie 
\begin{equation}\label{eq:GN}
{\| w\|}_{L^\infty(\R^d)} \lesssim {\| w\|}_{H^s(\R^d)} \lesssim \e^{-d/2}
{\| w\|}_{Y^s_\e}\, , \quad \text{for }s>d/2,
\end{equation} 
which we shall heavily use in the following.
\newpar
Multiplying the equation \eqref{eq:w} by $\overline{w^\e_N}$, integrating over $\R^d$, and taking the
imaginary part yields 
\begin{equation}\label{eq:nrj1}
\d_t {\left\| w^\e_N (t)\right\|}_{L^2} \lesssim  
{\left\| \, |\psi|^{2\sigma}\psi -
|v_N^\e|^{2\sigma}v_N^\e \right\|}_{L^2} +
\e^{N}{\left\|r^\e_N(t)\right\|}_{L^2},
\end{equation}
since $H^\e$ is self-adjoint and $|\kappa|=1$ by \eqref{nlcon}. 
To proceed further we recall the following Moser-type lemma, the proof of which is a straightforward 
generalization of those given in \cite{CMS, Ra}:
\begin{lemma}\label{lem:moser}
Let $R>0$, $s\in \N$, and $F(z)=|z|^{2\sigma}z$ for $\sigma\in\N$. Then there exists $C=C(R,s,\sigma,d)$ such that if $v$
satisfies
\begin{equation*}
{\left\| (\e x)^\alpha(\e\d)^\beta v \right\|}_{L^\infty(\R^d)} \leq R
\quad \text{for all }|\alpha|+|\beta|\leq s\, ,
\end{equation*}
and $w$ satisfies $\displaystyle {\left\| w \right\|}_{L^\infty(\R^d)} \leq
R$, then it holds 
\begin{equation*}
\sum_{|\alpha| +|\beta|\leq s}{\left\| (\e x)^\alpha (\e\d)^\beta \(F(v+w)
- F(v)\)\right\|}_{L^2(\R^d)} 
\leq C \sum_{|\alpha| +|\beta|\leq s}{\left\| (\e x)^\alpha (\e\d)^\beta
w\right\|}_{L^2(\R^d)}\, . 
\end{equation*}
\end{lemma}
We shall now use this lemma to factor out $w^\e_N$ in the right hand side of \eqref{eq:nrj1} and then taking advantage 
of the smallness of the remainder. By construction, $w^\e_N(0,x)=\O\(\e^{N+1}\)$ in any $Y^s_\e$. 
By Lemma~\ref{lemex1} we can find for fixed $\tau_0<\tau$ an $R>0$, such that if $N+1>d/2$, then 
\begin{equation}
\label{eq:solong}
{\left\|w_N^\e(t)\right\|}_{L^\infty}\leq R, 
\end{equation}
for $\e$ sufficiently small. Hence, as long as \eqref{eq:solong} holds, 
\eqref{eq:nrj1} and the above given Lemma~\ref{lem:moser}, with $s=0$, imply
\begin{equation*}
 \d_t {\left\| w^\e_N (t)\right\|}_{L^2} \leq C 
{\left\| w^\e_N (t) \right\|}_{L^2} +
C\e^{N}{\left\|r^\e_N(t)\right\|}_{L^2}.
\end{equation*}
Thus by a Gronwall type estimate, we get, as long as \eqref{eq:solong} holds, that 
\begin{equation}\label{eq:L2}
{\left\| w^\e_N (t)\right\|}_{L^2} \lesssim \e^{N}\, .
\end{equation}
for $t\leq \tau$. Next we shall show how to obtain similar estimates for the momenta and
derivatives of $w^\e_N$. Applying the operator $\e\nabla_x$ to \eqref{eq:w} yields (where, as before, $F(z):=|z|^{2\sigma} z$)
\begin{equation*}
\begin{aligned}
i\d_t (\e\nabla_x w^\e_N) =& \  H^\e (\e \nabla_x w^\e_N)  + 
\kappa \, \e\nabla_x \(F(\psi) - F(v_N^\e)\)\\
&+ \left[ \e\nabla_x,H^\e\right]w^\e_N - \e^{N+1}\nabla_x r^\e_N,
\end{aligned}
\end{equation*}
and hence we obtain the following energy estimate 
\begin{equation}
\begin{aligned}
\label{enes}
\d_t {\left\|\e\nabla_x w^\e_N(t)\right\|}_{L^2} \lesssim& \
{\left\|\e\nabla_x  
\(F(\psi)- F(\v_N^\e) \)\right\|}_{L^2} + {\left\| \left[
\e\nabla_x,H^\e\right]w^\e_N \right\|}_{L^2} \\
&+ \e^{N}{\left\|\e\nabla_x
r^\e_N\right\|}_{L^2}.  
\end{aligned}
\end{equation}
On the other hand we compute from \eqref{ham} that
\begin{equation*}
\left[\e\nabla_x,H^\e\right]  = \frac{1}{\e^2}\nabla_x V_\Gamma\left(\frac{x}{\e} \right) + \e
\nabla_x U(x).
\end{equation*}
Since $\nabla V_\Gamma$ is bounded and $\nabla U$ is
sub-linear, \eqref{enes} consequently yields 
\begin{equation*}
\begin{aligned}
\d_t {\left\|\e\nabla_x w^\e_N(t)\right\|}_{L^2} \lesssim & \  {\left\|
\e\nabla_x\(F(\psi) - F(v_N^\e)\) \right\|}_{L^2} + \frac{1}{\e^2} {\left\|
w^\e_N \right\|}_{L^2} + {\left\| \e x \, w^\e_N \right\|}_{L^2}\\
&+ \e^{N} {\left\|\e\nabla_x 
r^\e_N\right\|}_{L^2}.
\end{aligned} 
\end{equation*}
Thus, using again Lemma~\ref{lem:moser} (with $s=1$) together with Proposition~\ref{prop} and 
the estimate \eqref{eq:L2} we get
\begin{equation}\label{H1}
\begin{aligned}
\d_t {\left\|\e\nabla_x w^\e_N(t)\right\|}_{L^2} \lesssim \,  {\left\|\e\nabla_x w^\e_N 
 \right\|}_{L^2} + {\left\| \e x \,
w^\e_N \right\|}_{L^2}+ \e^{N-2}.
\end{aligned} 
\end{equation}
(Remark that the difference in the last term as compared to the semi-classical estimate 
obtained in \cite{CMS}.) To obtain an estimate for ${\left\| \e x \,  w^\e_N \right\|}_{L^2}$, we proceed analogously 
to obtain the following moment estimate:
\begin{equation}
\begin{aligned}
\label{moes}
\d_t {\left\|\e x\, w^\e_N(t)\right\|}_{L^2} \lesssim& \
{\left\|\e x  
\(F(\psi)- F(v_N^\e) \)\right\|}_{L^2} + {\left\| \left[
\e x,H^\e\right]w^\e_N \right\|}_{L^2} \\
&+ \e^{N}{\left\|\e x \,
r^\e_N\right\|}_{L^2}.  
\end{aligned}
\end{equation}
But, since $\left[\e x, H^\e\right]= -\e \nabla_x$ we get, as long as \eqref{eq:solong} holds
\begin{equation}\label{eq:x}
\begin{aligned}
\d_t {\left\|\e x \, w^\e_N(t)\right\|}_{L^2} &\lesssim  {\left\| \e x 
\(F(\psi) - F(v_N^\e)\) \right\|}_{L^2} + \left\| \e \nabla_x
w^\e_N \right\|_{L^2} + \e^{N}\left\|\e x \,r^\e_N\right\|_{L^2}\\
&\lesssim \left\| \e x \, w^\e_N(t)\right\|_{L^2} + \left\| \e\nabla_x
w^\e_N \right\|_{L^2} + \e^{N}.
\end{aligned} 
\end{equation}
Putting \eqref{H1} and \eqref{eq:x} together, we have
\begin{equation*}
\d_t \({\left\| \e\nabla_x w^\e_N \right\|}_{L^2}+ {\left\|\e x \, w^\e_N(t)\right\|}_{L^2}\) \lesssim  
{\left\| \e\nabla_x w^\e_N \right\|}_{L^2} + {\left\|\e x \, w^\e_N(t)\right\|}_{L^2}  + \e^{N-2}.
\end{equation*}
Hence a Gronwall lemma yields
\begin{equation}\label{X1}
{\left\| w^\e_N(t)\right\|}_{Y^1_\e}\lesssim \e^{N-2},
\end{equation}
as long as \eqref{eq:solong} holds and by induction one arrives at 
\begin{equation}
\label{Xk}
{\left\| w^\e_N(t)\right\|}_{Y^s_\e}\lesssim \e^{N-2s}\, .
\end{equation}
For $s>d/2$, and as long as \eqref{eq:solong} holds, the Gagliardo--Nirenberg type inequality \eqref{eq:GN} therefore implies
\begin{equation*}
{\left\| w^\e_N(t)\right\|}_{L^\infty(\R^d)}\lesssim \e^{-d/2}{\left\| w^\e_N(t)\right\|}_{Y^s_\e}\lesssim\e^{N-2s-d/2}.  
\end{equation*}
Hence if indeed $N-2s-d/2>0$ holds true, a continuity argument shows that
\eqref{eq:solong} is valid up to times $|t|= \tau$, provided $\e$ is sufficiently small. 
In particular, $w^\e_N$, and hence $\psi$, is well defined up to times $|t|= \tau_0<\tau$, for $0<\e\leq \e(\tau)$. 
\newpar
It remains to prove the estimate \eqref{eq:O}. Fix $s,N\in \N$ and let $s_1\geq s$ be such that 
$s_1>d/2$, as well as $N_1\geq 2s_1+N+1$. From \eqref{Xk} we conclude that 
\begin{equation*}
\sup_{t\in [-\tau_0, \tau_0]}
{\left\| w^\e_{N_1}(t)\right\|}_{Y^{s_1}_\e}\lesssim
\e^{N_1-2s_1}\lesssim \e^{N+1}.
\end{equation*}
Since $N_1>N$, it is therefore straightforward, that 
\begin{equation*}
\sup_{t\in [-\tau_0, \tau_0]}{\left\| v^\e_N(t)-
v^\e_{N_1}(t)\right\|}_{Y^{s_1}_\e}\lesssim \e^{N+1},
\end{equation*}
and hence, we deduce that \eqref{eq:O} holds for any $s,N\in\N$. 
\end{proof}
In the above given proof, the initial data $\psi_I$ is assumed to be well prepared 
up to any order $K\in \N$. Indeed this rather strong assumption can be relaxed as 
the next result will show. To this end we introduce the following notation:
\newpar
\textbf{Notation.} For every $\alpha \in \R$ we denote by $[\alpha]\in\N$, the \emph{ceiling of $\alpha$}, \ie 
the smallest integer which is larger than or equal to $\alpha$.
\begin{proposition}
\label{prop:finale}
Let $\widetilde \psi(t)$ be the solution of \eqref{snls1} corresponding to an initial data 
$\widetilde \psi_I$, which satisfies Assumption \ref{assw} for any $K\in \N$. On the other hand, let 
$\psi(t)$ be the solution corresponding to an initial data $\psi_I$, where $\psi_I$ is such that Assumptions~\ref{assw} is satisfied 
for $K\ge [3d/2]$. Then for any $\tau_0 \in
]-\tau,\tau[$, there exists $\e_0>0$ such that for $0<\e\leq
\e_0$, $\psi^\e(t)$ is defined up to times $|t| \leq \tau_0$ and moreover it holds:  
\begin{equation*}
\sup_{t\in [-\tau_0, \tau_0]} {\left\|\psi(t) - \widetilde
\psi(t)\right\|}_{Y^s_\e}  
= \O\(\e^{K+1 - 2s}\)\, ,\quad \text{for }s \ge 0\, .
\end{equation*} 
\end{proposition}
\begin{proof} Since the proof follows the lines of the one for Theorem~\ref{theo:stab}, we
shall be rather brief. Similarly as before we introduce 
$$\widetilde w (t,x):= \psi (t,x)- \widetilde \psi(t,x).$$ 
Then $\widetilde w(t)$ solves
\begin{equation*}
\left\{
\begin{aligned}
i\d_t \widetilde w &= H^\e\widetilde w  + \kappa \, \(|\psi|^{2\sigma}\psi - |\widetilde
\psi|^{2\sigma}\widetilde \psi\)\, , \\
\widetilde w\big|_{t=0} &=  \O\(\e^{K+1}\)\quad \text{in }Y^s_\e\text{ for any}s\in\N.
\end{aligned}
\right.
\end{equation*}
(Note that there is no remainder $r_N^\e$ in this case.) We can then argue as in the above given proof. 
We have that initially it holds
\begin{equation*}
{\| \widetilde w (0,\cdot) \|}_{L^\infty} \lesssim \, \e^{-d/2} {\|
\widetilde w (0,\cdot) \|}_{Y^s_\e}\lesssim \,  \e^{K+1-d/2}\, , \quad
\text{provided } s>\frac{d}{2}. 
\end{equation*}
With $K+1\geq [3d/2]+1>d/2$, the same arguments as in the proof of Theorem~\ref{theo:stab} yield 
\begin{equation*}
{ \| \widetilde w(t) \|}_{Y^s_\e}\lesssim \, \e^{K+1-2s},
\end{equation*}
as long as \eqref{eq:solong} holds. 
Since $K+1>d$, we can choose $s>d/2$ 
such that $K+1-2s> d/2$, \ie we can choose $s$ such that $K+1 > [d/2+2s]= [3d/2]$. Therefore the above given estimate and \eqref{eq:GN} show that
\eqref{eq:solong} 
holds up to times $|t|=\tau_0$, for $\e\ll 1$. 
\end{proof} 
Theorem~\ref{theo:stab} and Proposition~\ref{prop:finale} then finally lead to the following statement, proving also 
Theorem \ref{theo:typique}:
\begin{corollary}
If $\psi_I$ satisfies Assumption~\ref{assw} with $K\geq [3d/2]$, then there exists 
$\e_0>0$ such that for $0<\e\leq \e_0$, the solution $\psi(t)$ to \eqref{snls1} is defined on the 
time-interval $[-\tau_0,\tau_0 ]$, for any $\tau_0 < \tau$ and the following estimate holds:
\begin{equation}
\label{l2es}
\sup_{t\in [-\tau_0, \tau_0]}{\left\| \psi(t) - v_0^\e(t)\right\|}_{L^2(\R^d)} 
= \O\(\e\).
\end{equation} 
Additionally, if $K>[3d/2]$, then we have
\begin{equation}\label{lies}
\sup_{t\in [-\tau_0, \tau_0]}{\left\| \psi(t) - v_0^\e(t)\right\|}_{L^\infty(\R^d)} 
= \O\(\e\).
\end{equation} 
\end{corollary}
\begin{proof} From Proposition~\ref{prop:finale} we deduce that Theorem~\ref{theo:stab} holds with $K\geq [3d/2]$ 
The $L^2$ estimate \eqref{l2es} then is nothing but \eqref{eq:O} with $N=s=0$. The $L^\infty$ estimate
\eqref{lies} follows similarly from Theorem \ref{theo:stab} and \eqref{eq:GN}.
\end{proof}
In other words we deduce that the solution of \eqref{snls1} can be, up to an error of $\O(\e)$, approximated by the leading order 
asymptotic solution $v_0$, obtained from the multiple scales expansion, if the initial data is well prepared, \ie 
including correctors up to $K=[3d/2]$, which is slightly stronger than what was required for the semi-classical result 
given in \cite{CMS}. There the analogous condition for the correctors has been $K\geq d$. 
On the other hand one might guess that the leading order estimates \eqref{l2es}, \eqref{lies} are true 
even if the initial data is only ``correct'' 
up to leading order. The techniques used in the above given proofs though, do not allow the conclusion that this is indeed the case. 
Note however, that the higher order correctors in the initial data tend to zero, as $\e \rightarrow 0$, 
in practically every reasonable sense. 
\begin{remark} 
Finally, let us remark that one could also study the semi-classical asymptotic behavior of the homogenized 
NLS, \ie the limit $h\rightarrow 0$ of \eqref{effnls}, although, in view of the given scaling arguments and 
Remark \ref{screm}, the word ``semi-classical'' should rather be understood here in a purely mathematical sense. 
To this end, the well known WKB-type method derived in \cite{Gr} can be adapted under suitable conditions on $M^*$. 
In this case however, one can only hope for local-in-time results, \ie results up to caustics.
The combined limit $\e/h \rightarrow 0$, $h \rightarrow 0$ though, seems to be more subtle, in particular due to the 
somewhat hidden dependence of $\tau$ on $h$ in the above given results. 
\end{remark}
\newpar
\noindent {\bf Acknowledgments}. The author is grateful to R. Carles for helpful discussions on this work.


\bibliographystyle{amsplain}

\end{document}